\documentclass[preprint2,numberedappendix]{emulateapj-rtx4}
\usepackage{natbib}
\usepackage{amsmath}
\usepackage{graphicx,times,bm,url}
\usepackage{afterpage,lscape}
\graphicspath{{./fig/}{./png/}}

\newcommand{\EQ}{\begin{equation}}
\newcommand{\EN}{\end{equation}}
\newcommand{\EQA}{\begin{eqnarray}}
\newcommand{\ENA}{\end{eqnarray}}

\newcommand{\Fig}[1]{Figure~\ref{#1}}

{}
{}
{}

{}
{}
{}
{}
{}
{}
{}
{}
{}
{}
{}
{}
{}
{}
{}
{}
{}
{}
{}
{}
{}

{}
{}
{}

{}

{}
{}

\newcommand{\rr}{\mbox{\boldmath $r$} {}}

\newcommand{\kk}{\bm{k}}

\newcommand{\xx}{\bm{x}}

\newcommand{\uu}{\mbox{\boldmath $u$} {}}

\newcommand{\ee}{\mbox{\boldmath $e$} {}}
\newcommand{\nn}{\mbox{\boldmath $n$} {}}

\newcommand{\FF}{\mbox{\boldmath $F$} {}}

\newcommand{\nab}{\mbox{\boldmath $\nabla$} {}}

\newcommand{\dd}{{\rm d} {}}

\def\Rm{\mbox{\rm Re}_{\rm M}}

\def\Rey{\mbox{\rm Re}}

\def\km{k_{\rm m}}

\usepackage{color}

\usepackage{siunitx}

\begin{document}

\title{Estimating the rate of field line braiding in the solar corona by photospheric flows}

\author{S. Candelaresi}
\author{D. I. Pontin}
\affiliation{Division of Mathematics, University of Dundee, Dundee, DD1 4HN, United Kingdom}

\author{A. R. Yeates}
\affiliation{Department of Mathematical Sciences, Durham University,
Durham, DH1 3LE, United Kingdom}

\author{P. J. Bushby}
\affiliation{School of Mathematics, Statistics and Physics, Newcastle University, Newcastle upon Tyne, NE1 7RU, United Kingdom}

\author{G. Hornig}
\affiliation{Division of Mathematics, University of Dundee, Dundee, DD1 4HN, United Kingdom}

\begin{abstract}
In this paper we seek to understand the timescale on which the photospheric motions on the Sun braid
coronal magnetic field lines.
This is a crucial ingredient for determining the viability of the braiding
mechanism for explaining the high temperatures observed in the corona.
We study the topological complexity induced in the coronal magnetic field, primarily using plasma
motions extracted from magneto-convection simulations.
This topological complexity is quantified using the field line winding, finite time
topological entropy and passive scalar mixing.
With these measures we contrast mixing efficiencies of the magneto-convection simulation, a benchmark
flow known as a ``blinking vortex'', and finally photospheric flows inferred from sequences of observed
magnetograms using local correlation tracking.
While the highly resolved magneto-convection simulations induce a strong degree of field line winding
and finite time topological entropy, the values obtained from the observations from the plage region are
around an order of magnitude smaller.
This behavior is carried over to the finite time topological entropy.
Nevertheless, the results suggest that the photospheric motions induce complex tangling of the
coronal field on a timescale of hours.
\end{abstract}

\section{Introduction}

To understand the dynamics of solar plasmas it has become evident that we need to study
the topology of the magnetic field
lines, particularly in relation to reconnection and heating
\citep[e.g.\ ][]{Parker-1983-264-635-ApJ, Greene-1988-93-A8-JGRA, Lau-Finn-1990-350-672-ApJ,
Canfield1996c, Craig-Sneyd-2005-232-1-SolPhys, Pontin_etal11,
Yeates-Hornig-2011-18-102118-PhysPlasm, Guo-Ding-2013-779-2-ApJ, Low-2013-768-7-ApJ,
Park-Kusano-2013-778-1-ApJ, Cemeljic-Huang-2014-21-3-PhysPlasm, Brookhart-Stemo-2015-114-14-PRL,
Russell-Yeates-2015-22-3-PhysPlasm, Pontin-Candelaresi-2016-58-5-PlPhysContrFus,
Sun-Zhang-2016-830-1-ApJL}.
For topologically non-trivial magnetic field constructions in the form of
knots \citep{Ranada-Trueba-1995-202-337-PhysLetA, knotsDecay11, Smiet-Candelaresi-2015-115-5-PRL},
braids \citep{Yeates_Topology_2010, Wilmot-Smith-Pontin-2011-536-A67-AA}
and links \citep{fluxRings10}
we know that the magnetic helicity, which is a manifestation of linkage, knottedness
and braiding \citep{MoffattKnottedness1969}, imposes restrictions on the evolution of the field
\citep{Woltjer-1958-489-91-PNAS, Chandrasekhar-Woltjer-1958-44-4-PNAS, ArnoldHopf1974,
Taylor-1974-PrlE, Ricca2008RSPSA}.
This is particularly pronounced in the solar atmosphere, where, due to the high magnetic Reynolds
number, the helicity is conserved on dynamical time scales.
Braiding or twisting of the field lines can then play a critical role in the dynamics: this braiding/twisting can be
induced either below the solar surface before the flux emerges into the atmosphere, or after the flux emerges
in response to photospheric flows.

Explaining the huge temperature increase from the solar surface to the corona remains one of the most enduring
problems in solar physics (the ``coronal heating problem").
The energy that must be supplied to the corona to maintain its million-degree temperature
can be estimated by quantifying energy losses \citep{withbroe1977}.
A number of previous works have sought to quantify the energy injected into the coronal
field by examining photospheric flows.
Initial estimates were based on assessing `typical' motions of a twisting or
braiding type \citep{berger1993, zirker1993}.
More recently, Poynting flux estimates have been made based on velocity fields
extracted from observations.
\citet{Yeates-Hornig-2012-539-AA} and \citet{Welsch2015} used Fourier Local Correlation Tracking (FLCT) to estimate
photospheric velocities in a plage region.
Using these velocities they obtained a Poynting flux into the corona of around
$\SI{5e4}{\watt\per\meter\tothe{2}}$.
\cite{Shelyag2012} used magneto-convection simulations to study the vertical
Poynting flux, identifying a dominant contribution from plasma motions in strong
intergranular magnetic flux concentrations.

These above studies can give lower 
bounds on the energy injected into the corona, but do not
provide clues to the mechanism by which this energy might be dissipated.
Here we aim to assess specifically the braiding mechanism proposed by \cite{parker1972}.
A key ingredient of Parker's hypothesis is that once the magnetic field topology becomes ``sufficiently complex''
(i.e.\ the field lines become ``sufficiently tangled''), current sheets spontaneously develop leading to a rapid
conversion of magnetic energy to thermal energy. 
This hypothesis was recently put on firmer footing by \cite{pontin2015a}, who proved that any
force-free equilibrium with tangled magnetic field lines must contain thin current layers
\citep[the argument was extended to include fields close to force-free by][]{Pontin-Candelaresi-2016-58-5-PlPhysContrFus}.
Nevertheless, the efficiency of the braiding mechanism for heating the coronal plasma relies crucially
on the -- so far unknown -- timescales for the energy build-up
(the field line tangling) and energy release processes.

In this paper we study the efficiency by which the coronal field lines are tangled/braided in a topological sense 
by photospheric motions.
We principally address this by examining flows derived from an MHD simulation of solar convection.
These are then contrasted with the results from flows inferred from observations.
We quantify the field line tangling using various measures, including the passive scalar spectrum,
the finite time topological entropy and the winding number
\citep[e.g.\ ][]{Prior-Berger-2012-278-2-SolPh, Mangalam-Prasad-2017-AdvSpaSci}.
With these we aim to address the questions:
``How efficient are the motions on the photosphere in inducing twisting?''
and ``What time is necessary for the Sun to induce a complex entangled braid?''.

In the following we will first explain the details of our benchmark flow (section \ref{sec: blinking vortex})
and the numerical data from the convection simulation we use (section \ref{sec: magneto-convection}),
before we explain how we preprocess these data (section \ref{sec: Helmholtz Decomposition}) and analyze them in
section \ref{sec: passive scalar}.
We then go into the details of measuring entangling by using the winding number
and the finite time topological entropy in section \ref{sec: entanglement}
and conclude by comparing the simulation results with calculations using observational data
in section \ref{sec: observations}.

\section{Benchmark Flow: The Blinking Vortex}
\label{sec: blinking vortex}

\begin{figure}[t!]\begin{center}
\includegraphics[width=\columnwidth]{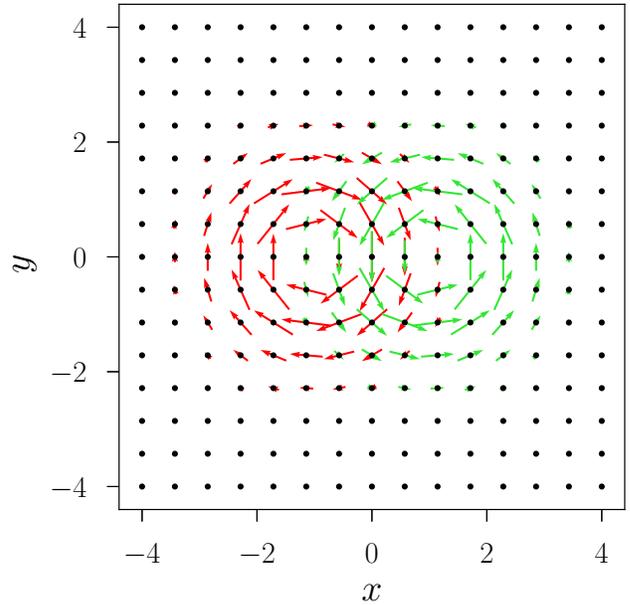}
\end{center}\caption[]{
Representation of the blinking vortex flow at two different times in red (left)
and green (right) arrows.
We switch periodically between the two driving vortices every $8$ time units.
This figure is available online as an animation.
}
\label{fig: blinking_vortex}
\end{figure}

Below we will discuss the tangling induced by the flows extracted from simulations that mimic the convective layer of the Sun.
In order to quantify the efficiency of that tangling we make use of a well-studied benchmark flow,
termed a ``blinking vortex" \citep{Aref-1984-143-1-JFluidMech}.
This flow comprises alternating positive and negative
vortices whose locations are off-set such that they only partially overlap
(see \Fig{fig: blinking_vortex}).
This motion is mostly confined within the domain $[-4, 4]\times[-3, 3]$ and alternates
every $8$ time units by smoothly switching them on and off (all units being non-dimensional).
Performing this motion for times between $0$ and $48$ results in
highly tangled fluid particle trajectories.
For more details about this flow see e.g.\ \cite{Candelaresi-Pontin-2017-27-9-Chaos},
equation (11).

When imposed as boundary footpoint motions, this flow generates a braided magnetic field with
a complex field line mapping, that has been studied in a series
of previous works \citep[e.g.\ ][]{Wilmot-Smith-Hornig-2009-696-1339-ApJ,
Wilmot-Smith-Hornig-2009-704-1288-ApJ, Yeates_Topology_2010, Wilmot-Smith-Pontin-2010-516-A5-AA,
Pontin_etal11, Wilmot-Smith-Pontin-2011-536-A67-AA, Russell-Yeates-2015-22-3-PhysPlasm, Ritchie-Wilmot-Smith-2016-824-19-ApJ}.
A sub-set of the field lines form a `pig-tail' braid and, due to the equal but opposite
alternating vortices, the net twist is zero.
Moreover, this pattern of opposing, overlapping twists has been shown to constitute a
maximally efficient protocol for generating small scales in the flow in the
analogous problem of fluid stirring with three stirring rods \citep{boyland2000}.
That is, our benchmark flow is highly efficient at inducing field line tangling.

\section{Magneto-Convection Simulations}
\label{sec: magneto-convection}

We consider three-dimensional MHD simulations of convectively-driven dynamos in a Cartesian domain
\citep[see][where more details can be found]{Bushby-Favier-2012-106-508-GAFD,Bushby-Favier-2014-562-A72-AA}.
Using the depth of the convective layer as the characteristic 
length scale in the system, the dimensions of the domain are given by $0\le x \le 10$, $0 \le y \le 10$ and $0\le z \le 1$,
with $z=0$ corresponding to the upper boundary.  
All quantities are assumed to be periodic in the horizontal ($x$ and $y$) directions.
The layer is heated from below and cooled from above, with 
the temperature fixed at each boundary. 
These upper and lower bounding surfaces are also assumed to be impermeable and stress-free, whilst the magnetic
field is constrained to be vertical at $z=0$ and $z=1$. 

The particular simulation that is considered here is a fully nonlinear dynamo calculation, 
with kinetic Reynolds number $\Rey \approx 150$ and magnetic Reynolds number $\Rm \approx 400$.
We focus upon the time-evolution of the velocity $\uu$ and vertical magnetic field $B_z$ at the top boundary ($z=0$),
extracting these quantities from $60$ snapshots, each of which has a horizontal resolution of $512^2$.
The time cadence of these snapshots is ca.\ $0.61$ in dimensionless units. One time unit corresponds to the time taken
for an isothermal sound wave to travel a distance of one unit across the surface of the domain.
However, it is perhaps more informative to note that this time cadence is approximately one fifth of the convective turnover time
\citep[which, based upon the rms velocity and the depth of the domain,
is ca.\ $3$ in these dimensionless units, see][]{Bushby-Favier-2012-106-508-GAFD}.
The convective turnover time is also comparable to the time taken for the (subsonic) convective motions to travel across 
a typical granular width of 2.5 time units. 
This granular timescale, which is arguably the most appropriate to consider when analyzing surface flows,
is the temporal normalization that is typically adopted below.

While the underlying simulations are fully three-dimensional we only make use of the
horizontal velocities at the top boundary of the domain (which plays the role of the photosphere in the model).
This invariably leads to apparent enhanced compression in areas of down-flows and
expansions in areas of up-flows.
Such compression effects lead to computational difficulties when we come to address field line tangling.
To circumvent these difficulties we decompose the velocity into divergent and rotational
(incompressible) parts, and make use of the latter for our calculations
(see section \ref{sec: Helmholtz Decomposition}).

\section{Helmholtz-Hodge Decomposition}
\label{sec: Helmholtz Decomposition}

For the horizontal velocity from the simulations (and later for flows inferred from observations)
we perform a Helmholtz-Hodge decomposition.
The 2d velocity is then written as the sum of three orthogonal terms:
\EQ\label{eq: Helmholtz decomposition}
\uu = \uu_{\rm i} + \uu_{\rm c} + \uu_{\rm h},
\EN
which are the incompressible, compressible and harmonic terms, respectively.
Here orthogonal means $\int\uu_{\rm i}\cdot\uu_{\rm c}\ \dd^2\xx = \int\uu_{\rm i}\cdot\uu_{\rm h}\ \dd^2\xx
= \int\uu_{\rm c}\cdot\uu_{\rm h}\ \dd^2\xx$.
They satisfy the conditions
\EQ
\uu_{\rm i} = \nab\times(\psi \ee_z), \qquad \uu_{\rm c} = \nab\phi, \qquad \uu_{\rm h} = \nab\chi,
\EN
with the differentiable scalar fields $\psi$, $\phi$ and $\chi$.
The velocities $\uu_{\rm i}$ and $\uu_{\rm c}$ satisfy the boundary conditions
$\uu_{\rm i}\cdot\nn = 0$ and $\uu_{\rm c}\cdot\nn = 0$, where $\nn$ is the normal vector to the boundary.

We compute the three flow components by solving the Poisson equation for the scalar fields while
respecting the given boundary conditions:
\EQA
\nab^2\psi = -\ee_z\cdot\nab\times\uu, \qquad \left.\psi\right|_{\partial V} = 0 \\
\nab^2\phi = \nab\cdot\uu, \qquad \left.\nn\cdot\nab\phi\right|_{\partial V} = 0 \\
\nab^2\chi = 0, \qquad \left.\nn\cdot\nab\chi\right|_{\partial V} = \nn\cdot\uu|_{\partial V}.
\ENA
For the calculations described below of the induced topological complexity, we use the incompressible part $\uu_{\rm i}$ only.
It is shown later that the contribution to the field line tangling of the flow component $\uu_{\rm c}$ is insignificant.

\section{Qualitative Mixing Patterns: Passive Scalar}
\label{sec: passive scalar}

A surface motion like the one simulated in turbulent convection leads to a mixing of the fluid.
Any features that are initially stretching over large length scales will then be mapped
into small scales, and \emph{vice versa}, at a rate that depends on how efficient the mixing is.
Allowing the field line footpoints to be transported
by these flows, and assuming an ideal evolution in the corona, we can make a direct association
between the tangling of these fluid particle
trajectories in time and the induced tangling of the anchored coronal magnetic flux tubes.

\begin{figure}[t!]\begin{center}
\includegraphics[width=1\columnwidth]{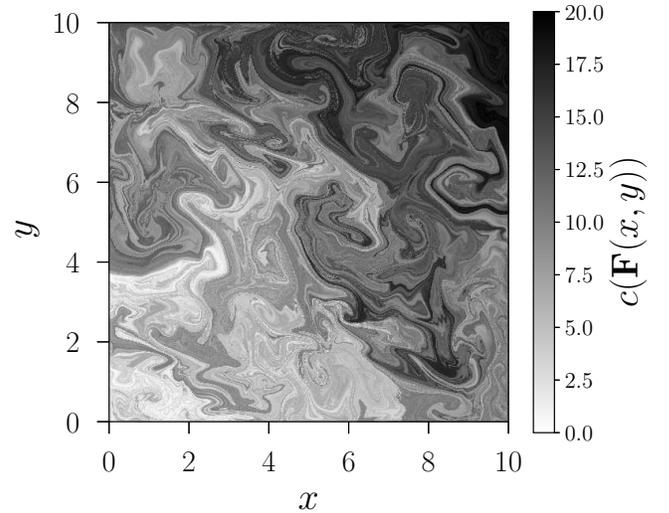}
\end{center}
\caption[]{
Passive scalar distribution after being mapped through the motions of the fluid
for the magneto-convection simulations at time $36$.
The initial distribution is $c(x, y) = x + y$.
This figure is available online as an animation.
}
\label{fig: passive_scalar}
\end{figure}

We first examine the complexity induced by the flows in a qualitative way.
By knowing the mapping of particles from positions $(x, y)$ to $\FF(x, y, t)$
at a time $t$, we can compute the mapping (pull-back) of a passive scalar (0-form or function)
$c(\xx)$, similar to \cite{Candelaresi-Pontin-2017-27-9-Chaos}.
This provides a visual representation of the mixing properties of the flow.
For our initial large-scale signal we use a simple gradient in $x$ and $y$ of the form
$c(x, y) = x + y$.
The mapped distribution, or equivalently the pull-back, is simply $c(\FF(x, y))$.
From the distribution of the mapped passive scalar (\Fig{fig: passive_scalar})
we can clearly see the turbulent nature of the thermo-convective simulations.
We observe fine-scale structures and a high level of fluid mixing.

A natural question is whether there exists any characteristic scale in the pattern obtained, or a rather a whole spectrum.
We cannot, however, judge from \Fig{fig: passive_scalar} if there are any particular length scales forming.
In order to seek the existence of characteristic length scales, we take the
two-dimensional Fourier transform
\EQ \label{eq: passive scalar fourier}
\mathcal{F}\{c(\FF(x, y))\}(\kk) = \int_{V} c(\FF(x, y)) e^{i\kk\cdot\xx}\ \dd x \ \dd y,
\EN
with the integration domain $V = [x_{\rm min}, x_{\rm max}]\times[y_{\rm min}, y_{\rm max}]$.
From this we compute the shell-integrated power spectrum of the passive scalar as
\EQ \label{eq: passive scalar shells}
\hat{c}(k) = \int\limits_{k-\delta k/2}^{k+\delta k/2} \mathcal{F}\{c(\FF(x, y))\}(\kk)\ \dd^2 k,
\EN
with the shell width $\delta k$.

\begin{figure}[t!]\begin{center}
\includegraphics[width=1\columnwidth]{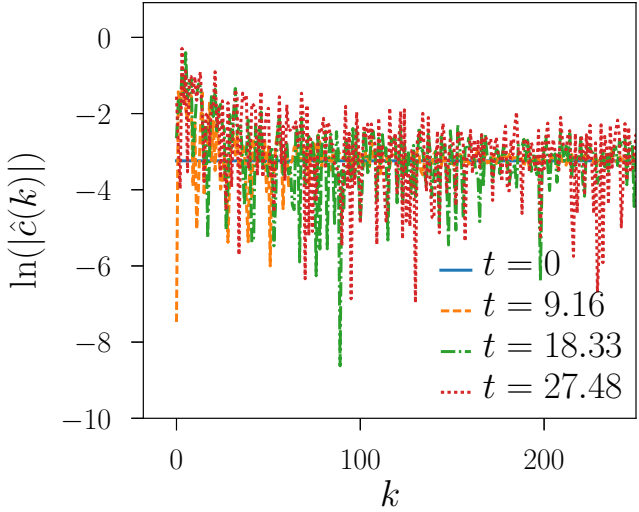}
\end{center}
\caption[]{
Fourier spectrum in time for the mapped passive scalar for the magneto-convection simulations at different times.
Here the time unit is in dimensionless code units.
}
\label{fig: fourier}
\end{figure}

Given that the simulation is turbulent we expect no such scales to arise.
This is indeed what we observe from the time dependent power spectrum of the
passive scalar distribution (see \Fig{fig: fourier}).
Already \cite{Candelaresi-Pontin-2017-27-9-Chaos}
showed that also for the blinking vortex benchmark flow small-scale structures form quickly,
but that in that case as well a characteristic scale is absent.

\section{Quantifying the Entanglement}
\label{sec: entanglement}

While the passive scalar transport and associated spectra give a visual impression
of the mixing, they do not encode any topological information about the trajectories.
In order to measure the induced braiding we consider two measures, one being the
(finite time) topological entropy \citep{Candelaresi-Pontin-2017-27-9-Chaos} that
is conceptually similar to the Lyapunov exponent, which was computed by
\cite{Favier-Bushby-2012-690-262-JFM} for closely related convective flows.
However, first we calculate the \emph{winding number} for field lines in the
domain \citep{Prior-Yeates-2014-787-2-ApJ, Mangalam-Prasad-2017-AdvSpaSci}.

\subsection{Winding Number}
\label{sec: winding}

For a magnetic field configuration it was shown \citep{Prior-Yeates-2014-787-2-ApJ} that the vertical magnetic
field weighted winding number corresponds to the magnetic field line helicity, which
encodes the topology of the field.
Let $\rr_1(t)$ and $\rr_2(t)$ be the trajectories of two particles with
starting positions $\rr_1(0) \ne \rr_2(0)$ on the photosphere.
Those trajectories are generated by the velocity field $\uu(\rr, t)$ with
\EQ
\frac{\dd \rr_1(t)}{\dd t} = \uu(\rr_1(t), t), \qquad \frac{\dd \rr_2(t)}{\dd t} = \uu(\rr_2(t), t).
\EN
The angle between the two particles at time $t$ is then calculated as
\EQ
\Theta(\rr_1(0), \rr_2(0), t) = \arctan\left(\frac{y_2(t) - y_1(t)}{x_2(t) - x_1(t)}\right),
\EN
and its time derivative is
\EQA
 && \frac{\dd \Theta(\rr_1(0), \rr_2(0), t)}{\dd t} =  \nonumber \\
 && \left( \uu(\rr_2(t), t) - \uu(\rr_1(t), t) \right)
\cdot \ee_z\times\frac{\rr_2(t) - \rr_1(t)}{|\rr_2(t) - \rr_1(t)|^2}.
\ENA
Summing over all trajectories $\rr_2$ gives us the net winding rate of all other trajectories around the trajectory $\rr_1$:
\EQ
w(\rr_1(0), t) = \int\limits_{(0, 0)}^{(L_x, L_y)} \frac{\dd \Theta(\rr_1(0), \rr_2(0), t)}{\dd t}\ \dd \rr_2(0).
\EN
By integrating over time and averaging over space we obtain the averaged net winding
around each trajectory, or averaged winding:

\EQ
\omega(\rr_1(0), T) = \frac{1}{L_x L_y} \int\limits_{0}^{T} w(\rr_1(0), t)\ \dd t,
\EN
with the final time $T$.

This number will depend on the integration time $T$, and we need to
circumvent this dependence when we make comparisons between different flows.
Furthermore, there is an arbitrariness in identifying time units in both
the simulations and the blinking vortex flow.
(A similar arbitrariness exists in the details of the time profile in the ``blinking vortex'' flow:
one could equally choose some other dependence that produces the same field line mapping
between the bottom and top boundaries.)
For these reasons we need to apply a normalization for the averaged winding $\omega$ that eliminates
such bias.

We choose to normalize by an averaged trajectory length in the $xy$-plane with respect to
the typical granular size $l_{\rm granules}$:

\EQ \label{eq: normalization winding}
q(T) = \frac{1}{l_{\rm granules}L_x L_y} \int\limits_{0}^{T} \int\limits_{(0, 0)}^{(L_x, L_y)} |\uu|\ \dd x \ \dd y \ \dd t.
\EN
This number $q$ is a measure of the average distance traveled by the trajectories
in terms of the granular size $l_{\rm granules}$ within the time $T$, while the time it takes to cross a
granule of size $l_{\rm granules}$ is $\tau_{\rm granules} = T/q$.
With this normalization we can write the normalized winding number as
\EQ \label{eq: normalized winding}
\Omega(\rr_1(0), T) = \frac{\omega(\rr_1(0), T)}{q(T)}.
\EN
For simplicity we will write $\Omega(\rr_1) = \Omega(\rr_1(0), T)$.
This quantity should be interpreted as follows.
It measures the average winding of trajectories around a particular trajectory $\rr_1$ during one granular crossing time.

For the granular size in the simulation data we use the value given by \cite{Bushby-Favier-2012-106-508-GAFD}
of a quarter of the box size, i.e.\ $2.5$ in code units. 
Since the blinking vortex set-up does not have a granular scale we use the diameter of one of the vortices as
a proxy for such a scale.
Here that size is $2$ units.
Furthermore, to take account of the fact that the flow does not fill the entire domain (see \Fig{fig: blinking_vortex}),
we do not average $|\uu|$ over the domain $\{x,y,t\}\in [-4, 4]\times[-4, 4]\times[0, 48]$,
but only over a region within which significant flows exist --  chosen to be $[0, 2]\times[-1, 1]\times[0, 48]$.

\begin{figure}[t!]\begin{center}
\includegraphics[width=1\columnwidth]{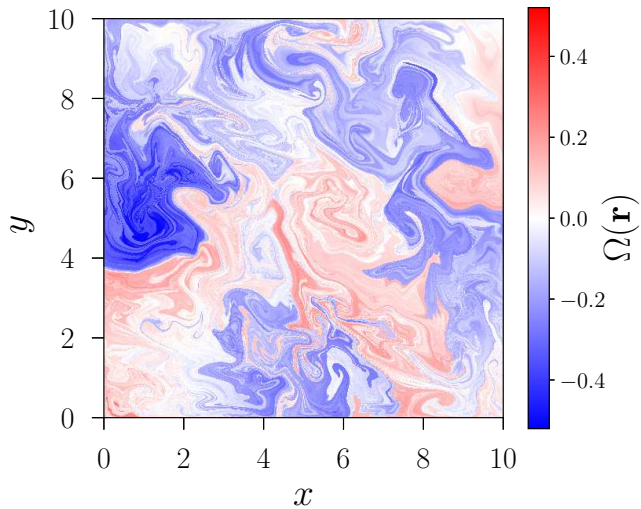}
\end{center}
\caption[]{
Winding number $\Omega(\rr)$ for trajectories starting at positions $\rr = (x, y)$ for the
magneto-convection simulation.
This figure is available online as an animation.
}
\label{fig: winding_Paul}
\end{figure}

\begin{figure}[t!]\begin{center}
\includegraphics[width=1\columnwidth]{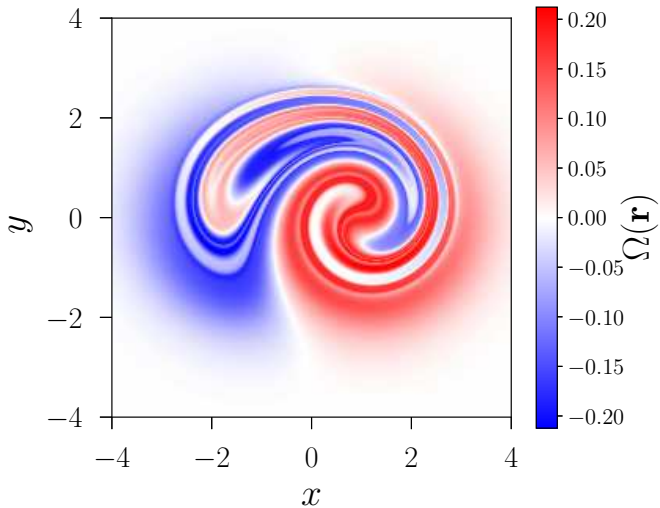}
\end{center}
\caption[]{
Winding number $\Omega(\rr)$ for trajectories starting at positions $\rr = (x, y)$ for the blinking vortex braid
at normalized time $t/\tau_{\rm granules} = 2.063$.
This figure is available online as an animation.
}
\label{fig: winding_e3_256_un0}
\end{figure}

The normalized, averaged winding over all trajectories in the domain is shown in \Fig{fig: winding_Paul}
for the magneto-convection simulations (at normalized time $t/\tau_{\rm granules} = 3.043$),
and in \Fig{fig: winding_e3_256_un0} for the blinking vortex
(at normalized time $t/\tau_{\rm granules} = 2.063$).
Both exhibit a complex pattern, although there exist much smaller scales in the pattern obtained from the simulations. 
Note that the times are only by chance this similar.
However, if we had chosen to use a longer or shorter time frame to perform our calculations
the final results would have changed only in terms of complexity with finer structures appearing in the plots,
while the extreme values converge quickly with time.

For a quantitative comparison of the tangling between the two flows we can consider the maximum or the spatial average
of the modulus of $\Omega$.
While both of these quantities depend on the integration time $T$, it is seen in the animated version of the Figures that
they converge as $T$ is increased.

Selecting representative times to make a comparison, we find a mean of the absolute value of $\Omega$ for the convective
simulations of $0.122$ and a maximum of $0.521$.
For the blinking vortex, we obtain a lower value of mean $|\Omega|$ (computed within the region $[-3, 3]\times[-2.5, 2.5]$,
corresponding to the region of significant twisting) of the absolute value of $0.0243$, and a maximum of $0.2122$.
This region is smaller than the region of significant velocities, as peripheral particles tend to induce
lower amount of twist, despite their significant velocities.
This is contrary to our expectations from the known highly efficient mixing property of this flow.
This is discussed further below.
In order to achieve a comparable winding number for the blinking vortex motion,
we would need to decrease $l_{\rm granules}$ relative to the size of the vortex pattern by a factor of
$0.434/0.2122 = 2.045$, leading to a granular size of $2/2.045 \approx 1$.

\begin{figure}[t!]\begin{center}
\includegraphics[width=1\columnwidth]{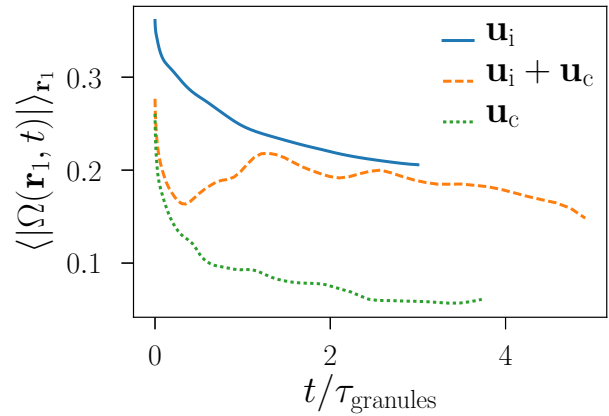}
\end{center}
\caption[]{
Spatially averaged absolute winding number as a function of  time for the incompressible part of the velocity
$\uu_{\rm i}$, the compressible part $\uu_{\rm c}$ and their sum.
}
\label{fig: winding_t_nonlinear_Re150_Rm400_inc_vc_Bz1kG_uf0_filterNone_un0_dt095}
\end{figure}

We finally comment on the effect of the decomposition described in Section \ref{sec: Helmholtz Decomposition}.
When splitting the velocity field into incompressible and compressible parts
using the Helmholtz-Hodge decomposition, we claimed that, for the winding number, contributions from $\uu_{\rm c}$
can be neglected.
We show this by computing $\langle |\Omega(\rr_1(0), t)| \rangle_{\rr_{1}}$ -- which is the spatial
average of the norm of $\Omega(\rr_1(0), t)$ -- for the two flow components $\uu_{\rm i}$ and $\uu_{\rm c}$ separately as well as their sum.
We clearly see that for the compressible part of the velocity $\uu_{\rm c}$ alone this number
quickly drops close to $0$ (see \Fig{fig: winding_t_nonlinear_Re150_Rm400_inc_vc_Bz1kG_uf0_filterNone_un0_dt095}),
while for $\uu_{\rm i}$ and $\uu_{\rm i} + \uu_{\rm c}$ it levels off at a finite value.
This justifies the usage of $\uu_{\rm i}$ alone when computing the winding number.

\subsection{FTTE}

The finite time topological entropy (FTTE) is a measure of the topological complexity of a flow
\citep{Adler-Konheim-1695-114-309-TrAmMathSoc, Sattari-Chen-2016-26-3-Chaos, Candelaresi-Pontin-2017-27-9-Chaos}.
It was shown by \cite{Newhouse-Pignataro-1993-72-5-JStatPhys} that it can be interpreted as the
exponential stretching rate of a material line $\gamma$ by the flow (specifically, the maximal stretching exponent
over all possible material lines $\gamma$ in the domain).
We can make use of this interpretation to measure the topological complexity  using the FTTE,
given a mapping $F(\gamma)$ of a material line $\gamma$.
Through the repeated application of the mapping $F$ we can compute the FTTE.
However, for the given velocity fields we cannot meaningfully construct such a repeated mapping,
since the flows are not time-periodic.
Therefore we make use of the mapping $F(\gamma, t)$ at time $t$ and compute the FTTE as
\EQ\label{eq: h}
h(F, \gamma, t) = \frac{1}{t} \ln\left(\frac{l(t)}{l_0} \right),
\EN
where $l_0$ is the length of the initial line $\gamma$, and $l(t)$ is the length of the mapped line, $F(\gamma,t)$.

In order to resolve the potentially highly entangled mapped line, we adaptively insert points
on the original line $\gamma$ where needed \citep[see][for details]{Candelaresi-Pontin-2017-27-9-Chaos}.
If we want to compare the FTTE of the blinking vortex \citep[investigated in detail by][]{Candelaresi-Pontin-2017-27-9-Chaos}
with the convective flows, we again need to normalize the time in some appropriate way.
Therefore, as before, we scale the time $t$ with the granule crossing time $\tau_{\rm granules}$ for both the simulations
and the blinking vortex.

\begin{figure}[t!]\begin{center}
\includegraphics[width=1\columnwidth]{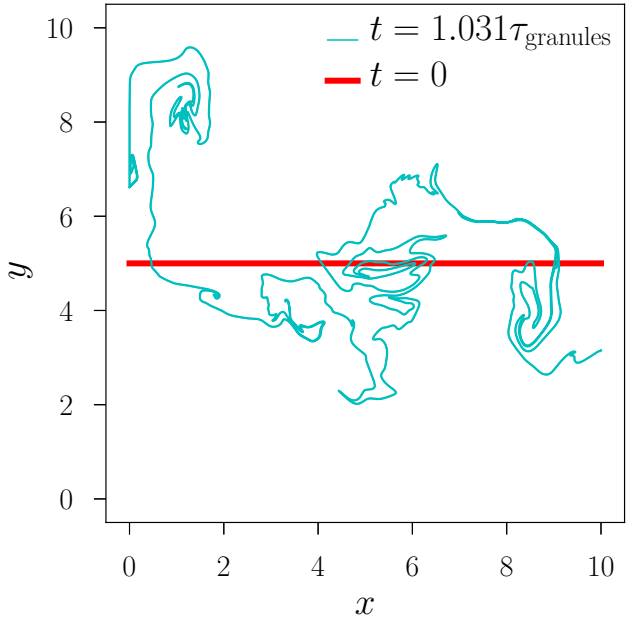}
\end{center}
\caption[]{
Mapped material line (cyan) for the magneto-convection simulation time $t = 1.031\tau_{\rm granules}$ compared
to its initial position (red).
This figure is available online as an animation.
}
\label{fig: mapping_nonlinear_Re150_Rm400_inc_Bz1kG_uf0_filterNone_un0_dt095}
\end{figure}

\begin{figure}[t!]\begin{center}
\includegraphics[width=1\columnwidth]{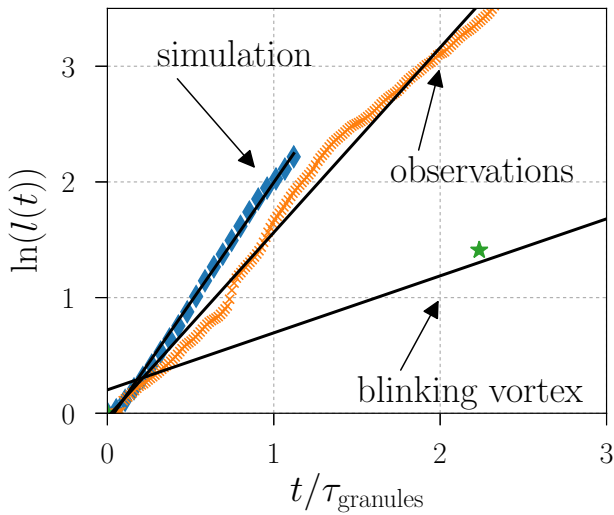}
\end{center}
\caption[]{
Logarithmic material line stretching in dependence of the normalized time $t/\tau_{\rm granules}$
together with linear fits for the magneto-convection simulations, the benchmark blinking vortex flow and
the observational data.
Note that the data points for the blinking vortex (stars) go well beyond the range of the plot and its linear
fit is performed on $13$ data points.
}
\label{fig: entropy_combined}
\end{figure}

We choose $\gamma$ to be the horizontal line centered in $y$ crossing the entire domain
in the $x$-direction.
(It is found that the choice of initial line makes little difference to the obtained value of $h$, due to the complexity of the flow.)
A mapping of this initial line is shown in \Fig{fig: mapping_nonlinear_Re150_Rm400_inc_Bz1kG_uf0_filterNone_un0_dt095}.
Due to the computational expense and the exponential growth of computing
time with increasing time, we compute the FTTE by using
only a small number of time steps from the simulations.
This turns out to be more than sufficient for estimating the entropy,
which requires only that we access the phase in which the material line
length grows exponentially.
As seen in \Fig{fig: entropy_combined}, a good fit is obtained even for the
short period of time over which we are able to compute, allowing for a relatively accurate
estimate to be obtained for $h$, c.f.\ equation \eqref{eq: h}.
For the magneto-convection simulations we use 20 time steps and
calculate a value for the FTTE of $h = 2.078$ (see \Fig{fig: entropy_combined}).

For the blinking vortex mapping we performed the calculations for the topological entropy
in a previous paper \citep{Candelaresi-Pontin-2017-27-9-Chaos}.
For comparison of notation, the blinking vortex mapping corresponds to one-third of the ``$E3$'' mapping
considered therein, and here we normalize the time by the granule crossing time.
We find that the blinking vortex flow exhibits a level of efficient mixing which
results in an FTTE of $0.4928$.
So, the FTTE for the simulations is higher by a factor of $4.217$.
This means that it takes 4.217 times longer to reach an equivalent state of entanglement in
the blinking vortex flow than in the flow from the simulations.
Similar to the winding number we attribute the higher value for the FTTE in the simulations to the fact that the turbulent
and space filling motions of the simulations induce a velocity pattern
that can be considered to be more chaotic.

\section{Comparison with Observations}
\label{sec: observations}

\subsection{Analysis of Flows Derived from Observations}

\begin{figure}[t!]\begin{center}
\includegraphics[width=1\columnwidth]{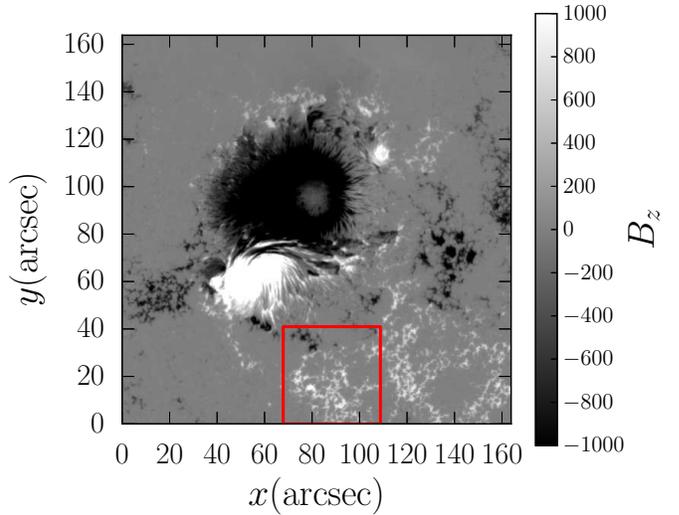}
\end{center}
\caption[]{
Magnetogram in units of Gauss for the vertical magnetic field at 12th of December
2006, 14:04 UT with the red area being used for our calculations.
This data is taken from \cite{Welsch-Kusano-2012-747-130}.
This figure is available online as an animation.
}
\label{fig: magnetogram_t0}
\end{figure}

We finally turn to compare the results obtained so far to those from observational data.
While  horizontal photospheric flows cannot be measured directly, they can be inferred by
various different methods, though each has its limitations \citep{Welsch-Abbett-2007-670-1434-ApJ}.
We use line-of-sight magnetogram data of the active region 10930 that is based
on Hinode/SOT observations \citep{Tsuneta-Ichimoto-2008-249-2-SolPhys} of Fe I at
\SI{6302}{\angstrom} 
\citep[see references in][for more details]{Yeates-Hornig-2012-539-AA}.
From these magnetograms, \cite{Fisher-Welsch-2008-383-373-ASPC} extracted velocity fields
using local correlation tracking \citep{November-Simon-1988-333-427-ApJ}.
With an observational noise level of \SI{17}{G} the tracking threshold for the magnetic features
was chosen to be \SI{15}{G} when extracting the velocities.
Subsequently the magnetograms were binned into blocks of $2\times 2$ pixels from a resolution of
\SI{0.16}{\arcsecond} to \SI{0.32}{\arcsecond} which corresponds to \SI{232.09}{\km}
(with \SI{1}{\arcsecond} on the Sun corresponding to \SI{725.281}{\km}).
The observations start at 14:00 UT on 12th December 2006 and run until 02:58 UT on 13th December 2006.
To reduce noise this time series is averaged over 4 images, which results into a cadence of \SI{121}{\second}.
An image of the magnetogram data is shown in \Fig{fig: magnetogram_t0}.

Observational data are inherently affected by noise.
This is already seen in the magnetic field.
For the computed velocity field from local correlation tracking the noise
is only increased, resulting in frequent spikes of over
\SI[per-mode=symbol]{1000}{\km\per\second}, while
for most of the domain the velocities remain below \SI[per-mode=symbol]{1}{\km\per\second} in magnitude.
This noise is greatly reduced by over two orders of magnitude once we
remove the purely divergent part of the velocity data using the Helmholtz-Hodge decomposition as described in section \ref{sec: Helmholtz Decomposition}.

\begin{figure}[t!]\begin{center}
\includegraphics[width=1\columnwidth]{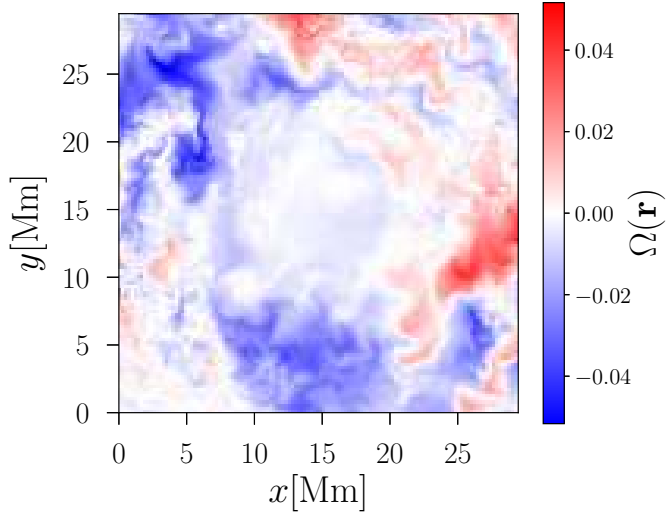}
\end{center}
\caption[]{
Winding number $\Omega(\rr)$ for trajectories starting at positions $\rr = (x, y)$ for the observed
velocity field using the ca.\ $12.8$ hours of data.
This figure is available online as an animation.
}
\label{fig: winding_obs_128_t367_inc_Bz1kG_uf0_filterNone_un0}
\end{figure}

From observations we know that the average granular size is of the order
of \SI{1}{\mega\meter} \citep[e.g.][]{Priest-2014-MHDSun}, which we use in our normalization for the winding number
which leads to $q_{\rm obs} = 2.72$ and $\tau_{\rm granules} = \SI{4.696}{\hour}$ 
using the definitions in Section \ref{sec: winding}.
For the observed velocity field we find maxima of the
normalized winding number $\Omega(\rr)$ of the order of $0.05$
(see \Fig{fig: winding_obs_128_t367_inc_Bz1kG_uf0_filterNone_un0}),
while for the mean of the magnitude we obtain $0.00994$.
This means that for every time the plasma travels (on average) a distance
of the granular size, the average winding angle of all field lines around a given field line increases/decreases
by $\SI{0.0517}{\radian} = \SI{2.959}{\degree}$.
The spatial distribution of this averaged, normalized winding $\Omega$ is surprisingly smooth.
With granules of the size of $4$ pixels (\SI{1}{\mega\meter}) one might expect variations
on the same scale.
However, we observe in \Fig{fig: winding_obs_128_t367_inc_Bz1kG_uf0_filterNone_un0} relatively
homogeneous patches with the same sign of winding number
over lengths of \SI{10}{\mega\meter}.

Again with an appropriate time normalization we can compute the FTTE for the observed flows for comparison with the previous results.
Doing this we obtain a value for the FTTE of $h = 1.598$ (see \Fig{fig: entropy_combined}).
We can now compute the physical time
for which the observed motions would lead to a braiding of an initially vertical magnetic field
equivalent to -- in the sense of having the same FTTE as -- the benchmark blinking vortex flow (with $l_{\rm granules}=2$).
The blinking vortex flow has a total logarithmic line stretching of $1.41$
which is obtained within a normalized time of $2.235$.
With an FTTE of $h = 1.598$ such a line stretching is obtained by the observations within
$0.682$ normalized times which corresponds to the physical time of
$0.682 \times \SI{4.485}{\hour} = \SI{3.059}{\hour}$.

\subsection{Magneto-Convection Degraded}

Comparing \Fig{fig: winding_Paul} and \Fig{fig: winding_obs_128_t367_inc_Bz1kG_uf0_filterNone_un0}
we observe significantly smaller lengths scales in the winding number plots for trajectories
from the simulated flows compared to those derived from observations.
In particular, the granular scales are now visible.
Perhaps more importantly, the peak winding number is a factor of 10 larger for the simulated flows.
One natural explanation for this could be the higher resolution of the simulated
flows compared to the observed flows.
To test this conjecture, we degrade the velocity data from the simulations such that granules
have the same resolution as in the observations.
This is done by applying a Butterworth filter
in $\kk$-space such that small-scale motions are filtered out.
For the simulation data we know that a typical granule extends ca.\
$2.5$ code units \citep{Bushby-Favier-2012-106-508-GAFD}, which translates
into $128$ pixels.
For the observational data we have a resolution of \SI{232.09}{\kilo\meter} per pixel
and a granular size of ca.\ \SI{1}{\mega\meter}, which means ca.\ $4$ pixels per granule.
Based on these numbers we degrade the simulation data by using
a Butterworth filter that filters out motions below the size of $32\times 32$ pixels.
In $\kk$-space this translates to a value of $16$, since the resolution is $512$ ($512/32$).
We can then calculate the winding number using this degraded velocity data.

\begin{figure}[t!]\begin{center}
\includegraphics[width=1\columnwidth]{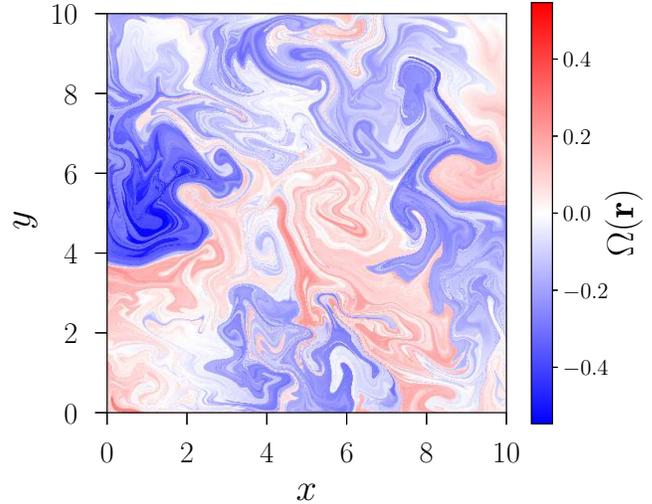}
\end{center}
\caption[]{
Winding number $\Omega(\rr)$ for trajectories starting at positions $\rr = (x, y)$ for the
magneto-convection simulation with the Fourier-filtered velocity.
This figure is available online as an animation.
}
\label{fig: winding_Paul butter}
\end{figure}

Recomputing the winding number for these degraded flows we find that, contrary to the above conjecture,
the values of $\Omega$ are similar to those computed
for the highly resolved original velocity data (see \Fig{fig: winding_Paul butter}).
These values are significantly higher than those obtained from the observational data.
We conclude that the observational resolution alone is not sufficient to explain
low values for the winding number and one needs to take into account the effects
from the velocity extraction method, as was suggested by
\cite{Welsch-Abbett-2007-670-1434-ApJ}.
Note in particular that in that study -- where a rate of helicity injection was compared
between exact values and those obtained by using FLCT to infer the velocity field --
the FLCT method was shown to under-estimate the helicity injection by a factor of $\sim 10$,
consistent with what we observe here (albeit for different underlying flow fields). 
Similar to the non-degraded calculations we compute a mean value for $|\Omega|$ of $0.128$.
In summary, the observed results using FLCT should be treated as inconclusive by themselves.

\section{Conclusions}

We have quantitatively and qualitatively investigated the efficiency of photospheric motions
in inducing non-trivial magnetic field line topology into coronal magnetic loops.
The efficiency of this tangling is crucial for evaluating the contribution of the braiding
mechanism first proposed by \cite{parker1972} for explaining the high temperature of the solar corona.
Recently Parker's hypothesis was put on firmer footing by \cite{pontin2015a}, who proved that any
force-free equilibrium with tangled magnetic field lines must contain thin current layers
\citep[the argument was extended to include fields close to force-free by][]{Pontin-Candelaresi-2016-58-5-PlPhysContrFus}.
They estimated that for coronal plasma parameters, significant energy release should be expected
after 3-6 iterations of the blinking vortex pattern
(where the mapping shown in Figure \ref{fig: winding_e3_256_un0} with $T=48$ corresponds to
three iterations).
Crucial to determining the efficiency of the braiding mechanism, then, are the timescales for
injecting such tangling and for the associated energy release.
Herein we have investigated the complexity induced in the coronal field by photospheric flows,
and we find that (even assuming a trivial identity mapping of minimal complexity at $t=0$)
such levels of complexity can be induced in a matter of hours.

For the magneto-convection simulations we saw a relatively high degree of
trajectory winding compared to the benchmark case of the blinking vortex flow.
This is reflected in the calculation of the FTTE that shows a similar difference.
We attribute this surprisingly high degree of winding to the presence of volume
filling turbulent motions.
Those lead to highly tangled particle trajectories which we observe in our calculations.

We also compared the results to observed velocity fields of a solar plage region
where the velocities had been extracted using the Fourier correlation tracking
method.
The winding number and the FTTE were significantly below the simulations and the
blinking vortex benchmark.
This cannot be due to the lower resolution alone (factor $32$ difference).
Tests with degrading the simulation data show similar values as the high resolution simulation data.
So it must be attributed to other effects, like the method of extracting
velocity information from a time series of magnetograms.
\cite{Welsch-Abbett-2007-670-1434-ApJ} compared different
methods for extracting velocity data from photospheric
magnetic field time series and found that the local correlation tracking approach
used here leads to a factor $10$ inferior magnetic helicity injection rate,
consistent with the factor we find for the winding number.
On the other hand, other approaches for velocity extraction tend to require vector magnetogram
data and/or further assumptions about the magnetic field structure.
A further potential source of the discrepancy is the different regions on the surface that we consider.
For the observed field we limit our calculations to a plage region that is relatively quiet.
By comparing with the pig-tail braid we also calculated that the plage region
induces as much mixing as the pig-tail braid in just \SI{3.059}{\hour}.

Our calculations shed some light into the Parker braiding problem for coronal magnetic fields.
It is clear that the build up of braids and tangles leads to small-scale variations in magnetic fields
that can further induce strong electric currents, reconnection and possibly heating.
From our calculations we show that this mechanism is feasible due to the presence of
turbulent motions that effect the tangling of the magnetic field on a timescale of hours.

\section*{Acknowledgements}
SC, DIP, ARY and GH
acknowledge financial support from the UK's
STFC (grants number ST/N000714 and ST/N000781).
PB would like to acknowledge the support of the Leverhulme Trust (grant number RPG-2014-427).
For the graphs we made use of the package {\sc Matplotlib} \citep{Hunter:2007}.

\bibliography{references}
\end{document}